\documentclass{ws-p8-50x6-00}

\begin{document}

\title{A Submillimeter Selected Quasar in the Field of Abell 478}  

\author{K.K.\ Knudsen, P.P.\ van der Werf and W.\ Jaffe}

\address{Leiden Observatory, P.O.\ Box 9513, NL--2300 RA Leiden, The Netherlands
\\E-mail: kraiberg@strw.leidenuniv.nl}


\maketitle

\abstracts{
We report the discovery of a $z=2.83$ quasar in the field of the 
cooling flow galaxy cluster Abell 478. This quasar was first detected 
in a submm survey of star forming galaxies at high redshifts, as the 
brightest source.  We discuss the optical spectrum and far--IR 
spectral energy distribution (SED) of this object. 
}

\section{Introduction}

We are performing a survey of gravitationally lensed submm 
sources --- a project aimed at studying the cosmic star formation history.  
Our survey covers 11 clusters of galaxies and has so far resulted 
in the detection of at least 25 submm sources. 
Here we present results on the brightest submm source detected.

\section{Observations and Results}

\begin{figure}
\begin{center}
\epsfxsize=17pc % will enlarge or reduce the postscript figures based on the xsize
 \epsfbox{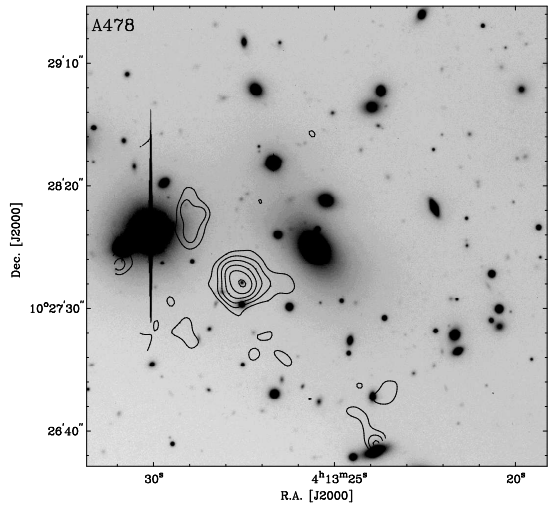} % postscript image file name
 \caption{The $I$--image of A478 overlayed with the contours of the SCUBA 
 $850\ \mu{\rm m}$ data.  The contours each represent a step of 4 mJy.  
 SMMJ04135+1027 is the very bright SCUBA source located SE of the center
 of the cluster.  The other fainter possible submm source is located NE of 
 the quasar and nearby the very bright star. 
  \label{fig:A478}}
\end{center}
\end{figure}

In the field of Abell478 we have detected with SCUBA (JCMT, Hawaii) a very 
bright submm source, $F_{850\mu{\rm m}} = 25\pm3\ {\rm mJy}$ and 
$F_{450\mu{\rm m}} = 63\pm20\ {\rm mJy}$, and possibly
a fainter source of $F_{850\mu{\rm m}} = 9 \pm 2\ {\rm mJy}$. 
The errors contain the detection uncertainty and the calibration error. 
To identify the sources we have obtained a deep $I$--band image with
FORS1 (VLT, Paranal). Our optical identification of the bright submm 
source, which we henceforth refer to as SMMJ04135+1027, 
is an $I = 20.5\pm 0.1\ {\rm mag}$ point source. There are no 
obvious optical counterparts for the faint submm source.  
In Fig.\ \ref{fig:A478} the $I$--band image is shown overlayed by the 
SCUBA contours. 

The Galactic extinction of A478 is $E(B-V) = 0.52\ {\rm mag}$ (Schlegel 
et al 1998). Hence, the corrected $I$ magnitude is $19.5\pm 0.1\ {\rm mag}$
assuming a Milky Way type extinction law with $R_V = 3.1$. 
All the optical observations presented in the rest of this paper have been 
corrected accordingly. 

\begin{figure}
\epsfxsize=20pc
\epsfysize=14.1pc
 \epsfbox{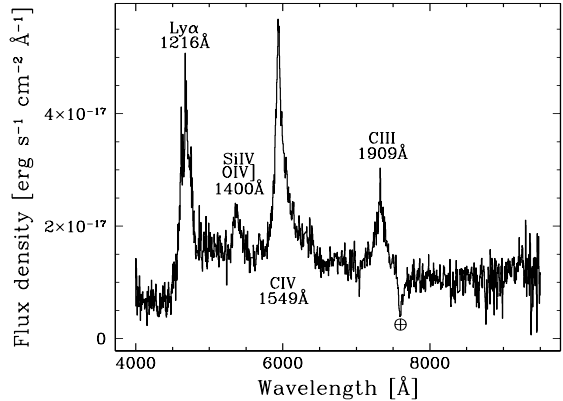}
 \caption{The spectrum of SMMJ04135+1027, corrected for Galactic 
 extinction.  \label{fig:spectrum} }
\end{figure}

To determine the redshift and the nature of SMMJ04135+1027 we
have obtained an optical spectrum also with FORS1. The spectrum was
observed in MOS mode with resolution $R=150$.  The spectrum exhibits the 
characteristica of high redshift quasars, such as broad emission lines
(e.g.\ Ly$\alpha$, C{\small IV}, C{\small III} and Si{\small IV}), 
Lyman Forest absorption bluewards of the Ly$\alpha$ emission line, and 
a power law continuum.  The redshift has been measured using the 
C{\small III} line to be $2.83\pm0.01$, which is consistent with the
other emission lines. 
The spectrum is shown in Fig.\ \ref{fig:spectrum}. 
If omitting the correction for Galactic extinction, the quasar 
appears very red. When corrected, as mentioned above, the quasar has 
a colour similar to optically selected quasars (Francis et al.\ 1992). 

\section{Discussion}

\subsection{Optical spectrum}

The spectrum shown in Fig.\ \ref{fig:spectrum} shows a number of unusual
features when compared to the QSO sample of Francis et al.\ (1992). 
In the first place, Ly$\alpha$ is strongly suppressed. The origin of 
the suppression may be continuum absorption by dust, or Ly$\alpha$ 
absorption associated with the QSO.  A higher resolution spectrum is 
needed to distinguish between these possibilities.  
Secondly, the C{\small IV} emission line is
remarkably strong with an equivalent width of $110\ {\rm {\AA}}$.

\subsection{The Far--IR SED}

The fluxes quoted here have not been corrected for the weak magnification
arising from the gravitational lensing caused by the intervening 
galaxy cluster.  In this case, no arcs are detected in the vicinity 
of the quasar, hence the gravitational magnification is expected to be
small and achromatic.

We have plotted the two SCUBA data points together 
with four known far--IR SEDs (Klaas et al 1999, Haas et al 1998, 
Sodroski et al 1997). The four SEDs have been redshifted to the
redshift of the quasar and scaled to the quasar flux at $\lambda_{obs} = 
850 \ \mu{\rm m}$.  We see that the data points correspond well to the 
SEDs of the two starburst galaxies, Arp220 and the Antennae, and to the 
low--z QSO PG0050+124. The cool Milky Way type of dust can easily be 
ruled out.  Shorter wavelength data would be needed to see if the hot
dust component that is found in low--z QSOs such as PG0050+124 is also
present in SMMJ04135+1027.  For an Arp220 SED, the total luminosity 
of SMMJ04135+1027 is $3\times 10^{13} {\rm L}_{\odot}$.

\begin{figure}
\epsfxsize=20pc
\epsfysize=14.5pc
 \epsfbox{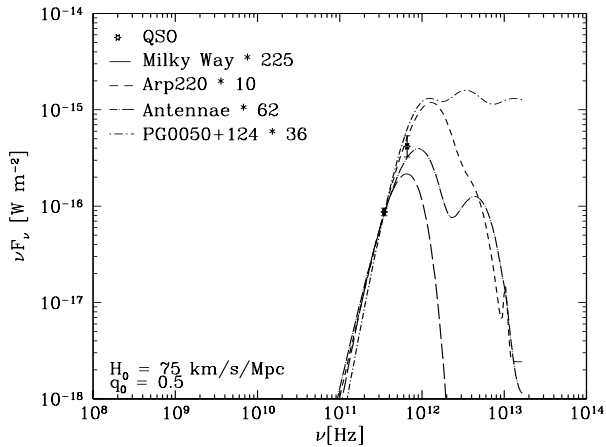}
\caption{The $850\ \mu{\rm m}$ and $450\ \mu{\rm m}$ data points overlayed
with the redshifted far--IR SEDs of the Milky Way, Arp220, the Antennae and 
PG0050+124.  \label{fig:SEDs} }
\end{figure}

\subsection{Contribution of AGNs to the submm samples}

A number of well-studied submm sources exhibit AGN features in their 
optical spectra (e.g.\ Ivison et al.\ 1998).  These are also typically the 
brightest submm sources in the samples. 
However, the fraction of AGNs in 
submm samples is still a debated issue.  Optical spectroscopy of the 
fainter submm sources has turned out to be very challenging,  therefore
it will be difficult to fully characterize the submm population with 
optical spectroscopy alone.  As shown by Fig.\ \ref{fig:SEDs}, shorter 
wavelength IR data may be used to find QSO--like dust emission. The most
promising method for finding AGNs in submm samples is by hard X--ray
observations. 
A recent study with {\em Chandra} and SCUBA 
of the relation between the resolved X-ray background and the resolved
FIR background finds almost no overlap between the two populations, 
which suggest that SCUBA sources primarily are powered by starburst 
(Fabian et al.\ 2000).  
This result then suggests that the fainter submm sources contain 
relatively few AGNs, and that the AGNs, that \underline{are} present 
in the submm samples, are to be found at the bright submm flux levels.

\end{document}